\documentclass[twocolumn]{aastex631}
\usepackage{graphicx}
\usepackage{amssymb}
\usepackage{amsmath}
\usepackage{epstopdf}
\usepackage{calc}
\begin{document}


\shorttitle{MeerKAT Atlas of Southern RBGS Galaxies}
\shortauthors{Condon et al.}

\pdfsuppresswarningpagegroup=1
\maxdeadcycles=1000

\title{A MeerKAT 1.28\,GHz Atlas of Southern Sources in the
  \emph{IRAS} Revised Bright Galaxy Sample}

\correspondingauthor{James J.~Condon} \email{jcondon@nrao.edu}

\author[0000-0003-4724-1939]{J.~J.~Condon}
  \affiliation{National Radio Astronomy Observatory, 520
    Edgemont Road, Charlottesville, VA 22903, USA}

\author[0000-0001-7363-6489]{W.~D.~Cotton}
  \affiliation{National Radio Astronomy Observatory, 520
    Edgemont Road, Charlottesville, VA 22903, USA}

\author[0000-0002-4939-734X]{T.~Jarrett}
\affiliation{Astronomy Department,
University of Cape Town,
Private Bag X3,
Rondebosch 7701,
South Africa}

\author[0000-0003-3948-7621]{L.~Marchetti}
\affiliation{Astronomy Department,
University of Cape Town,
Private Bag X3,
Rondebosch 7701,
South Africa}
\affiliation{INAF - Istituto di Radioastronomia, via Gobetti 101,
40129 Bologna, Italy}

\author[0000-0002-6479-6242]{A.~M.~Matthews}
\affiliation{Department of Astronomy, University of
Virginia, Charlottesville, VA 22904, USA}
\affiliation{National Radio Astronomy Observatory, 520 Edgemont Road,
Charlottesville, VA 22903, USA}

\author[0000-0003-2716-9589]{T.~Mauch}
\affiliation{South African Radio Astronomy Observatory (SARAO), 2 Fir
Street, Black River Park, Observatory, 7925, South Africa}

\author[0000-0001-5519-0620]{M.~E.~Moloko}
\affiliation{Astronomy Department,
University of Cape Town,
Private Bag X3,
Rondebosch 7701,
South Africa}

\begin{abstract}
  The {\it IRAS} Revised Bright Galaxy Sample (RBGS) comprises
  galaxies and unresolved mergers stronger than $S =
  5.24\,\mathrm{Jy}$ at $\lambda = 60\,\mu\mathrm{m}$ with galactic
  latitudes $\vert b \vert > 5^\circ$.  Nearly all are dusty
  star-forming galaxies whose radio continuum and far-infrared
  luminosities are proportional to their current rates of star
  formation.  We used the MeerKAT array of 64 dishes to make $5 \times
  3\,\mathrm{min}$ snapshot observations at $\nu =
  1.28\,\mathrm{GHz}$ covering all 298 southern (J2000 $\delta <
  0^\circ$) RBGS sources identified with external galaxies.  The
  resulting images have $\theta \approx 7\,\farcs5$ FHWM resolution
   and rms fluctuations $\sigma \approx
  20\,\mu\mathrm{Jy\,beam}^{-1} \approx 0.26\,\mathrm{K}$ low enough
  to reveal even faint disk emission. The rms position
  uncertainties are $\sigma_\alpha \approx \sigma_\delta \approx
  1\arcsec$ relative to accurate near-infrared positions, and the
    image dynamic ranges  are DR $\gtrsim 10^4:1$.  Cropped
  MeerKAT images of all 298 southern RBGS sources are available in
  FITS format from \url
  {https://doi.org/10.48479/dnt7-6q05}.
\end{abstract}




\section{Introduction}

The {\it IRAS} \citep{neu84} Revised Bright Galaxy Sample
\citep[RBGS,][]{san03} comprises all 629 {\it IRAS} sources
(individual galaxies or unresolved mergers) with $S(60\,\mu\mathrm{m})
\geq 5.24\, \mathrm{Jy}$ in the extragalactic sky defined by Galactic
latitude $\vert b \vert > 5^\circ$.  It is the far-infrared (FIR)
counterpart of the Revised Third Cambridge catalog of radio sources
with $S \geq 10\,\mathrm{Jy}$ at 178\,MHz \citep[3CR,][]{lai83}.
These FIR and radio flux-limited samples of strong sources differ in
that most of the $\lambda = 60\,\mu\mathrm{m}$ emission is from dust
heated by short-lived ($\tau \lesssim 100\,\mathrm{Myr}$) stars in
nearby (median redshift $\langle z \rangle = 0.01$) star-forming
galaxies (SFGs), while most of the radio sources are powered by active
galactic nuclei (AGNs) in distant ($\langle z \rangle \sim 1$) radio
galaxies and quasars.  However, the very tight FIR/radio correlation
of nearby SFGs \citep{hel85} makes radio continuum emission an
excellent complementary tracer of recent star formation, and the RBGS
selects the brightest SFG needles in the haystack of all radio
sources.

The radio continuum emission from SFGs is a combination of synchrotron
radiation from electrons accelerated in the supernova remnants of
massive ($M > 8 M_\odot$) very short-lived ($\tau < 30\,\mathrm{Myr}$)
stars and free-free emission from \ion{H}{2} regions ionized by the
most massive of those stars \citep{con92}.  At decimeter wavelengths,
the steep-spectrum synchrotron emission is much stronger than the
flat-spectrum free-free emission, so most SFGs have spectral indices
$\alpha \equiv d \ln (S) / d \ln (\nu) \approx -0.7$  \citep{con92}.

Combining FIR and radio observations of RBGS galaxies adds value to
both.  Nearly all SFGs are  optically thin to $\lambda =
  60\,\mu\mathrm{m}$ dust emission, so the {\it IRAS} RBGS is the
purest complete sample of star-forming galaxies in the local universe.
The majority of RBGS galaxies are also bright optically, but  a
  significant fraction of optically luminous galaxies have low
star-formation rates (SFRs) \citep{con19} and galaxies with the
highest SFRs are often too dusty to stand out in photographic
catalogs.  Aperture-synthesis radio observations can yield images with
signficantly higher angular resolution and smaller position errors,
and unusually low FIR/radio flux-density ratios can reveal the
presence of obscured AGNs  \citep{con19}.  The radio and FIR
morphologies of SFGs are similar, but the radio emission is blurred by
cosmic-ray diffusion \citep{mur08} and can reveal winds and
outflows from galactic disks \citep{ade13}.

The most important new population in the {\it IRAS} RBGS is the
subsample of $\approx 200$ luminous infrared galaxies (LIRGs) defined
by $L_\mathrm{IR}(8-1000\,\mu\mathrm{m}) \geq 10^{11} L_\odot$, where
$H_0 = 70\,\mathrm{km\,s}^{-1}\,\mathrm{Mpc}^{-1}$ was used to
calculate $L_\mathrm{IR}$ and $L_\odot \equiv 3.83 \times 10^{26}
\,\mathrm{W}$ is the nominal bolometric luminosity of the Sun.  A
  significant fraction of low-redshift LIRGs are merging systems
containing nuclear starbursts and/or AGNs \citep{san96}, and
LIRGs at redshifts $z \gtrsim 1$ produced most of the stars in the
universe today.  The Great Observatories All-sky LIRG Survey
\citep[GOALS,][]{arm09} is a comprehensive imaging and spectroscopic
survey of RBGS LIRGs.

The ``characteristic'' infrared luminosity $L^*_\mathrm{IR} \approx
10^{10.5}L_\odot$ \citep{san03} of RBGS galaxies is a large fraction
of their characteristic bolometric luminosity.  In contrast, radio
continuum emission is only an energetically insignificant tracer---the
typical IR/radio luminosity ratios of SFGs are $> 10^5$---making it
vulnerable to contamination by potentially more luminous AGN emission.

To quantify the FIR/radio flux density ratio, \citet{hel85} defined
the parameter
\begin{equation}\label{egn:qdef}
  q \equiv \log\Biggl[
    \frac{F_\mathrm{FIR} / (3.75 \times 10^{12}\,\mathrm{Hz})}
    {S\mathrm{(1.4\,GHz)}} \Biggr]\,,
\end{equation}
where $S\mathrm{(1.4\,GHz)}$ is the 1.4\,GHz flux density in
W\,m$^{-2}$\,Hz$^{-1}$ and $F_\mathrm{FIR}$ is the far-infrared flux
in W\,m$^{-2}$ in the band of width $3.75 \times 10^{12}\,\mathrm{Hz}$
between $\lambda = 42.5\,\mu\mathrm{m}$ and $\lambda =
122.5\,\mu\mathrm{m}$ calculated from
\begin{equation}\label{eqn:FIRdef}
  F_\mathrm{FIR} \equiv
  1.26 \times 10^{-14}
  [2.58 S(60\,\mu\mathrm{m}) + S(100\,\mu\mathrm{m})]
\end{equation}
using infrared flux densities $S(60\,\mu\mathrm{m})$ and
$S(100\,\mu\mathrm{m})$ in Jy.  Typical values for SFGs are
$S(100\,\mu\mathrm{m}) \approx 2 S(60\,\mu\mathrm{m})$ and $q = 2.3$,
making most RBGS galaxies with $S_{60\,\mu\mathrm{m}} \geq
5.24\,\mathrm{Jy}$ radio sources with $S(1.4\,\mathrm{GHz}) \gtrsim
40\,\mathrm{mJy}$.  The radio luminosities of RBGS galaxies with low
star-formation rates are not diluted by cold ``cirrus'' dust emission
in the diffuse interstellar medium heated by stars older than 30\,Myr.
In addition to being only a tracer, the synchrotron luminosity of an
SFG depends on too many unmeasured parameters to be modeled
quantitatively, so the theoretical foundation for the tight empirical
FIR/radio correlation is weak.

The main drawback of the {\it IRAS} images is their limited angular
resolution---half of the RBGS sources are unresolved by the $1\farcm5
\times 4\farcm5$ response of {\it IRAS} \citep{neu84} detectors at
$\lambda = 60\,\mu\mathrm{m}$, and the smallest galaxies that show
structural features at {\it IRAS} resolution have blue isophotal
angular diameters $\gtrsim 8'$ \citep{ric88}. Only at radio
wavelengths is the sensitivity of multi-element interferometers nearly
immune to the quantum noise which adds $T \approx 50\,\mathrm{K}
\times (\nu / \mathrm{THz})$ to the noise temperature of coherent
amplifiers, allowing large radio arrays such as MeerKAT to achieve
much higher angular resolution and astrometric accuracy than {\it
  IRAS}.  Comparisons of higher-resolution {\it Spitzer} $\lambda =
70\,\mu\mathrm{m}$ images of the nearest ($D <11.5\,\mathrm{Mpc}$)
galaxies with 1.4\,GHz radio images show similar FIR and radio
morphologies, and their small differences can usually be explained by
cosmic-ray diffusion smoothing the radio images and by cirrus emission
in the FIR images of quiescent SFGs \citep{mur08}.  Moreover, the
mid-infrared ``warm'' dust has been shown to be a good tracer of the
cold FIR \citep[e.g.,][]{clu17}, which enables using higher
resolution imaging from {\it Spitzer} and {\it WISE} that closer
matches that of the radio interferometry.

This paper presents new MeerKAT images with FWHM resolution $\theta
\approx 7\,\farcs5$ and rms fluctuations $\sigma \approx
20\,\mu\mathrm{Jy\,beam}^{-1}$ covering all 298 RBGS sources
identified with external galaxies in the southern hemisphere (J2000
$\delta < 0$).  The observations and imaging processes are described
in Section~\ref{sec:obs}.  The resulting MeerKAT Atlas of 1.28\,GHz
continuum images is presented in Section~\ref{sec:atlas}.
Table~\ref{tab:radio} lists the 1.28\,GHz total flux densities of
southern RBGS sources plus fitted source-componentparameters (peak and
integrated flux densities, deconvolved angular sizes, and J2000
positions).  Section~\ref{sec:summary} summarizes the MeerKAT Atlas
results and their significance.

The original RBGS catalog \citep{san03} used $H_0 =
75\,\mathrm{km\,s}^{-1} \mathrm{\,Mpc}^{-1}$ and $\Omega_\mathrm{m} =
0.3$ to calculate absolute quantities in a $\Lambda$CDM universe.  For
comparison with the more frequently used $H_0 =
70\,\mathrm{km\,s}^{-1} \mathrm{\,Mpc}^{-1}$, the distances in
Table~\ref{tab:iras} should be multiplied by $1.071$ and
$\log(L/L_\odot)$ should be increased by $0.060$.  This paper uses the
spectral-index sign convention $\alpha \equiv + d \ln S / d \ln \nu$.
Most SFGs have spectral indices near $\alpha = -0.7$, so
$S(1.4\,\mathrm{GHz}) \approx 0.94\,S(1.28\,\mathrm{GHz})$.

\section{MeerKAT Observations and Imaging} \label{sec:obs}

Our observations were scheduled as two MeerKAT projects: (1) the
director's discretionary time project DDT-20200520-TM-01 with six
observing runs in 2020 May and June plus (2) the ``open time'' project
SCI-20210212-TJ-01 with eight runs in 2021 February and March.  The 14
observing runs are labeled ``A'' through ``N'' and listed in
Table~\ref{tab:obs} along with their starting dates, UTC time ranges,
and calibration sources.  Each run was $\sim$8 hr long and cycled
through five 3~min integrations on each of $\gtrsim 20$ RBGS
targets passing near transit.  The DDT project pioneered the use of
``snapshot'' observations with MeerKAT to observe large numbers of
sources quickly, as described in the SARAO science commissioning
report ``Snapshot Observations with MeerKAT'' \citep{mau20b}.  Single
pointing positions of the $\Theta_{1/2} = 68\arcmin$ FWHM MeerKAT primary beam
were able to cover each of these five target groups:
 F02069$-$1022, F02071$-$1023, and F02072$-$1025; 09432$-$1405
and F09433$-$1408; 14214$-$1629 and 14216$-$1632; F14376$-$0004 and
F14383$-$0006; and F23156$-$4238, F23161$-$4230, and F23165$-$4231.

A single complex gain (both amplitude and phase) calibrator was
observed twice per hour throughout each run. In addition, the
flux-density/bandpass calibrators J0408$-$6545 and/or PKS~B1934$-$638
= J1939$-$6342 were observed with several 10~min scans in each run.
Our flux-density scale is based on the \citet{rey94} spectrum of
PKS~B1934$-$638 = J1939$-$6342:
\begin{eqnarray}
  \log(S) = & -30.7667 + 26.4908 (\log\nu)
  - 7.0977 (\log \nu)^2 \nonumber \\
   & +0.605334 (\log\nu)^3\, , \qquad\qquad\qquad\qquad\qquad
\end{eqnarray}
where $S$ is the flux density in Jy and $\nu$ is the frequency in
MHz. A single 10~min scan of the polarization calibrator 3C~138 or
3C~286 was made when possible, and the resulting polarization
calibration was applied to runs lacking scans on these polarization
calibrators as described in Section~\ref{sec:calib}.

The observed frequency range 856 to 1712 MHz was divided into 4096
spectral channels and the integration time was 8~s to minimize
bandwidth and time smearing.  Weak, irregular receiver response
  and strongly elliptical primary beams at the correlator band edges
limit the useful frequency range to $880 \lesssim \nu(\mathrm{MHz})
\lesssim 1670$ whose midpoint is $\nu \approx 1280\,\mathrm{MHz}$.
All combinations of the linearly polarized feeds were correlated and
recorded.  Typically, more than 60 of the 64 MeerKAT antennas were
operating during our observations.

\begin{deluxetable*}{c l c l}
  \label{tab:obs}
  \tabletypesize{\small}
\tablecaption{MeerKAT Observation Log}  
\centering
\tablehead{
  \colhead{Run} & \colhead{Start Date} & \colhead{UTC Time Range} & \colhead{Calibration Sources} }
\startdata
A &2020 May  29 & 19:47--27:45 & 3C~286, J1830$-$3602, J1939$-$6342 \\
B &2020 Jun. 11 & 14:46--22:44 & J0408$-$6545, 3C~286, 1424$-$4913, J1939$-$6342 \\
C &2020 Jun. 13 & 02:08--10:00 & J0408$-$6545, J0440$-$4333 \\
D &2020 Jun. 15 & 23:31--31:25 & 3C~138, J1939$-$6342 \\
E &2020 Jun. 20 & 04:48--14:44 & J0408$-$6545, J0616$-$3456 \\
F &2020 Jun. 21 & 12:01--20:00 & J0408$-$6545, J1154$-$3505, 3C~286 \\
G &2021 Feb. 20 & 00:33--07:35 & 3C~296, J1939$-$6342 \\
H &2021 Feb. 21 & 09:03--17:05 & J0408$-$6545, 3C~138, J1939$-$6342 \\
I &2021 Feb. 24 & 14:38--22:29 & J0408$-$6545, J0609$-$1542, 3C~138 \\
J &2021 Feb. 24 & 22:32--30:41 & J0408$-$6545, J1311$-$2216, 3C~286, J1939$-$6342\\
K &2021 Feb. 28 & 09:08--16:10 & J0408$-$6545, 3C~138, J1939$-6$342 \\
L &2021 Feb. 28 & 21:57--29:29 & J0408$-$6545, J1154$-$3505, 3C~286, J1939$-$6342 \\
M &2021 Mar. 05 & 11:35--19:25 & J0240$-$2309, J0408$-$6545, 3C~138 \\
N &2021 Mar. 23 & 17:02--24:40 & J0408$-$6545, J1120$-$2508, 3C~286, J1939$-$6342 
\enddata
\end{deluxetable*}

\subsection{Calibration}
\label{sec:calib}
The 144 spectral channels on each end of the bandpass were deleted and
the remaining channels were divided into 8 spectral windows for
calibration purposes.  Initial flagging and calibration were as
described in \citet{mau20} and \citet{cot20} using the {\it Obit}
\citep{cot08} software package.  Channels containing persistent,
strong interference were flagged.  Parallel-hand calibration consisted
of intermixed calibration and editing steps: after a first pass, the
flagging was kept, the calibration tables were deleted, and the
process was repeated.  Calibration included group delay, bandpass, and
gain calibration.  The bandpass calibrators J0408$-$6455 and
PKS~B1934$-$638 = J1939$-$6342 are so weakly polarized that they can
be used to fix the gain ratios of the two orthogonally polarized
feeds.  Gain calibration subsequent to bandpass calibration used only
Stokes $I$ to preserve those gain ratios.

The polarization calibration for all runs included instrumental
cross--hand delay and phase calibration based on the ``DelayCal''
calibration performed prior to the beginning of each observing run.
Runs that included a polarization calibrator had instrumental
polarization and residual cross-hand phases determined and applied to
the data.  The initial instrumental cross-hand delay calibration
leaves sufficiently stable residuals that the calibration derived from
a run with a polarization calibrator can be applied to another run
without one.

\subsection{Imaging Process}
All pointings were imaged in Stokes $I$, $Q$, $U$, and $V$ by the {\it
  Obit} task MFImage \citep{cot18}.  MFImage divides the sky into
small flat facets to approximate the curvature of the sky and
divides the data into spectral bins narrow enough that variations of
antenna gain and sky brightness with frequency do not lower image
quality.  The spectral bins were imaged independently and CLEANed
jointly.

Each pointing was fully imaged out to the first zero of the MeerKAT
primary beam $72\arcmin$ from the pointing center. Outlying facets
were added up to $90\arcmin$ from the pointing center to cover
individual sources from the NVSS \citep{con98} or SUMSS catalogs
\citep{mau03} having flux densities $S >1\,\mathrm{mJy}$ after
attenuation by the primary beam.  The frequency bins have 0.05 maximum
fractional bandwidth and the baseline-dependent time averaging was
constrained to neither exceed 30~s nor lower any amplitude in the
field of view by more than 1\%.  An {\it AIPS/Obit} Robust factor $R =
-1.5$ was used to produce nearly circular Gaussian restoring beams
with FWHM major and minor diameters $\theta_\mathrm{M} \approx
\theta_\mathrm{m} \approx 7\,\farcs5$.  Because the Rayleigh-Jeans brightness
temperature corresponding to a peak flux density $S_\mathrm{P}$ is
\begin{eqnarray}
  \Biggl( \frac {T_\mathrm{b}} {\mathrm{K}} \Biggr) \approx 0.26
  \Biggl( \frac {S_\mathrm{P}} {20\,\mu\mathrm{Jy\,beam}^{-1}} \Biggr)
  \times \qquad \nonumber \\
    \Biggl( \frac {7\,\farcs5} {\theta_\mathrm{M}} \Biggr)
    \Biggl( \frac {7\,\farcs5} {\theta_\mathrm{m}} \Biggr)
    \Biggl( \frac {1.28\,\mathrm{GHz}} {\nu} \Biggr)^2~,
\end{eqnarray}
an rms image fluctuation $\sigma_\mathrm{n} \approx 20\,\mu\mathrm{Jy\,beam}^{-1}$
is sufficient for detecting most optically selected SFGs, which have
median face-on disk brightness temperature $\langle T_\mathrm{b}
\rangle \gtrsim 1 \,\mathrm{K}$ at $\nu = 1.28\,\mathrm{GHz}$
\citep{hum81}.  At the median angular-size distance $D_\mathrm{A} =
D_\mathrm{C} / (1+z) \approx 50\,\mathrm{Mpc}$ of RBGS galaxies
\citep{san03}, $7\,\farcs5 = 1.8\,\mathrm{kpc}$.

Each run used a single complex-gain calibrator for target fields
separated by up to a radian on the sky, so the external phase
calibration is not always very accurate.  In order to mitigate this, a
relatively complicated self-calibration scheme was used.  All
pointings use three cycles of 30~s phase-only self calibration
preceded by CLEANs reaching residual peak flux densities of 10, 2, and
0.5 mJy\,beam$^{-1}$.  This helps prevent artifacts from the initial
calibration from corrupting the self-calibration model.  For fields
with peak flux densities in excess of $1\,\mathrm{Jy\,beam}^{-1}$, a
scan-averaged amplitude and phase self calibration was applied.
Sources with peaks in excess of $0.15 \,\mathrm{Jy\,beam}^{-1}$ were
centered in their own facets to facilitate CLEANing and improve the
image dynamic range.

CLEAN windowing used the MFImage ``autoWindow'' facility.  Final
CLEANing in Stokes $I$ with a loop gain of 0.05 proceeded to a
residual peak flux density $S_\mathrm{p} = 100\,\mu
\mathrm{Jy\,beam}^{-1}$ or a maximum of 50,000 components.  Stokes $Q$
and $U$ images were CLEANed to a depth of $50\,\mu
\mathrm{Jy\,beam}^{-1}$ or a maximum of 8,000 components, and Stokes
$V$ CLEANing was stopped after reaching
$50\,\mu\mathrm{Jy\,beam}^{-1}$ or a maximum of 1,000 components.

\subsection{Imaging Data Products}

Two of the 14 frequency planes were totally blanked to eliminate
  strong radio-frequency interference. MFImage produced CLEAN
  restored images in the 12 remaining frequency planes. Each image
  plane was divided by the primary beam attenuation pattern as
  described in \citet{mau20} to yield ``on sky'' brightnesses.  A
flux density and then a flux-density plus spectral index were fitted in
each pixel. If the higher-order fit did not increase $\chi^2$ by more
than 1.5, the spectral index was kept; otherwise, a default spectral
index $\alpha = -0.6$ was used.   The limited frequency range does not
allow usefully accurate spectral curvature fits.  The region
within $20\arcmin$ of each
target galaxy was extracted for further analysis.

\begin{figure}[!htb]
  \centering
  \includegraphics[width=0.5\textwidth,trim = {1.5cm 12.5cm 4.cm 3.5cm},clip]
  {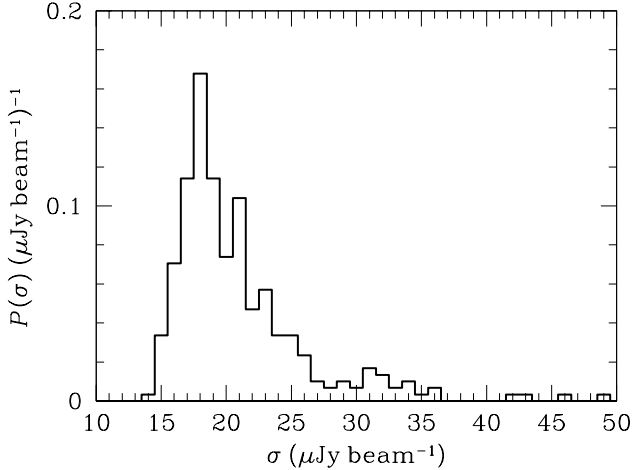}
  \caption{This histogram shows the probability distribution
    $P(\sigma)$ of rms brightness fluctuations $\sigma$ in the central
    $16\arcmin \times 16\arcmin$ squares of our MeerKAT RBGS images.
    The long tail is caused by sidelobes from target nuclei or
    background AGNs brighter than $S_\mathrm{p} \sim
    200\,\mathrm{mJy\,beam}^{-1}$.
  \label{fig:RBGSrms}}
\end{figure}

Off-source brightness fluctuations in the Stokes $I$ images are
produced by a combination of thermal noise, confusion by faint sources
overlapping within the synthesized beams, and residual sidelobes
of imperfectly calibrated strong sources.  We estimated the rms
brightness fluctuation in the central $16\arcmin \times 16\arcmin$
square of each image using the {\it AIPS} task IMEAN, which fits
a Gaussian to the peak of the pixel brightness distribution.  The rms
$\sigma$ of that Gaussian fit excludes the long tail of very bright
pixels on individual sources.  Figure~\ref{fig:RBGSrms} shows the
normalized distribution of $\sigma$ for our MeerKAT images.  The rms
thermal noise is $\sigma_\mathrm{n} \approx 15\,\mu
\mathrm{Jy\,beam}^{-1}$ and the rms confusion caused by very
  faint sources is $\sigma_\mathrm{c} \approx
  2\,\mu\mathrm{Jy\,beam}^{-1}$. The quietest images have $\sigma
  \approx \sigma_\mathrm{n}$ and the typical image has $\sigma
  \lesssim 20\,\mu\mathrm{Jy\,beam}^{-1}$.  Significantly larger
  values of $\sigma$ occur in fields containing residual
    sidelobes of strong ($S \gtrsim 200\,\mathrm{mJy}$) target or
  background sources.  The snapshot image dynamic range (ratio of
image peak flux density to $\sigma$) is typically between $10^4:1$ and
$3 \times 10^4:1$.

The final step in constructing the Atlas Stokes $I$ images was
correcting for the negative ``bowl'' that results from incomplete
CLEANing of an interferometric image lacking zero-spacing data.  We
measured the median brightness in a large source-free region
surrounding each target and subtracted that median brightness from the
image.  Most bowl depths were only $-1$ or
$-2\,\mu\mathrm{Jy\,beam}^{-1}$ and there are $\approx 57$ beam solid
angles per arcmin$^2$, so the corrections are small and add $\lesssim
0.1\,\mathrm{mJy}$ to the total flux densities of most target
galaxies.

Linear polarization was analyzed as described in \citet{cot20} using a
simple rotation-measure (RM) analysis to derive the intrinsic angle of
the polarized emission.  RMs and electric-vector position angles
corrected to zero wavelength were derived from a search in RM space.
The RM giving the highest polarized intensity $P = (Q^2 + U^2)^{1/2}$
was taken to be the actual RM, the highest $P$ as the polarized
intensity, and the polarization angle of the RM-corrected $Q+iU$ as
the electric-vector position angle corrected to zero wavelength.
Stokes $Q$ and $U$ images were extracted in the same regions as Stokes
$I$.  The rms image noise in the Stokes $Q$ and $U$ images is
  close to the thermal noise, $\sigma_\mathrm{n} \approx 15
  \,\mu\mathrm{Jy\,beam}^{-1}$.

Cropped MeerKAT images in FITS format covering all 298 southern RBGS
sources are available via
\url{https://doi.org/10.48479/dnt7-6q05} as a single 221 Mbyte
compressed file, and individual images are available as $\sim 1$ Mbyte
uncompressed FITS files.

\section{The MeerKAT 1.28\,GHz Atlas of Southern RBGS Sources}
\label{sec:atlas}

\subsection{{\it IRAS} data}

There are 300 RBGS sources south of J2000 declination $\delta =
0^\circ$.  We observed all but the two Magellanic Clouds because they
are only satellites of our Galaxy.  Our Galaxy and its satellites are
not external galaxies representative of the universe as a whole,
unbiased by our location.  The remaining 298 are listed in
Table~\ref{tab:iras}, which is a subset of \citet[table~1]{san03}.
There is one row per {\it IRAS} source listing: \\

{\it Column (1)}.---Our infrared source index number $N$ running from
1 through 298 for easier comparison with other tables and lists in
this paper.

{\it Column (2)}.---The most common galaxy name(s).  

{\it Column (3)}.---The {\it IRAS} source name. 

{\it Columns (4)--(5)}.---The {\it IRAS} J2000 right ascension $\alpha$ and
declination $\delta$. 

{\it Column (6)}.---RBGS flux density at $\lambda = 12\,\mu\mathrm{m}$ (Jy). 

{\it Column (7)}.---RBGS flux density at $\lambda = 25\,\mu\mathrm{m}$ (Jy). 

{\it Column (8)}.---RBGS flux density at $\lambda = 60\,\mu\mathrm{m}$ (Jy).

{\it Column (9)}.---RBGS flux density at $\lambda = 100\,\mu\mathrm{m}$ (Jy).

{\it Column (10)}.---Comoving distance $D_\mathrm{C}$ (Mpc) calculated
by \citet{san03} for a flat universe with $H_0 =
75\,\mathrm{km\,s}^{-1} \,\mathrm{Mpc}^{-1}$.  For $H_0 =
70\,\mathrm{km\,s}^{-1}\,\mathrm{Mpc}^{-1}$, this listed
$D_\mathrm{C}$ should by multiplied by $75/70 \approx 1.07$.

{\it Column (11)}.---Logarithm of the absolute IR [$8 <
  \lambda(\mu\mathrm{m}) < 1000$] luminosity $L_\mathrm{IR}/L_\odot$
in units of the solar bolometric luminosity $L_\odot = 3.83 \times
10^{26}\,\mathrm{W}$.  The infrared fluxes $F_\mathrm{IR}$ in
W\,m$^{-2}$ were extrapolated from {\it IRAS} flux densities in Jy
using \citep{san96}
\begin{eqnarray}
  F_\mathrm{IR} = 1.8 \times 10^{-14}
  (13.48 S_{12\,\mu\mathrm{m}} + 5.16 S_{25\,\mu\mathrm{m}} \nonumber \\
  + 2.58 S_{60\,\mu\mathrm{m}} + S_{100\,\mu\mathrm{m}})\,. \qquad\qquad\quad
\end{eqnarray}
The corresponding luminosities are $L_\mathrm{IR} = 4 \pi
D_\mathrm{L}^2 F_\mathrm{IR}$, where $D_\mathrm{L} = (1+z) D_\mathrm{C}$
is the bolometric luminosity distance.

{\it Column (12)}.---An asterisk indicates the source is one of the 94
southern sources in the GOALS \citep{arm09} sample of LIRGs defined by
$\log(L_\mathrm{IR}/L_\odot) \geq 11.0$ calculated using $H_0 =
70\,\mathrm{km\,s}^{-1}\,\mathrm{Mpc}^{-1}$ instead of $H_0 =
75\,\mathrm{km\,s}^{-1}\,\mathrm{Mpc}^{-1}$.

\startlongtable
\begin{deluxetable*}{C l l C c R R R R R C c}
  \centerwidetable
  \label{tab:iras}
  \tabletypesize{\small}
  \centering
  \tablecaption{The Southern RBGS Sample: {\it IRAS} Data}  
\tablehead{
  \colhead{\llap{In}d\rlap{ex}} & \colhead{Common} & \colhead{{\it IRAS}} &
  \colhead{$\alpha$} & \colhead{$\delta$} &
  $S_{12\,\mu\mathrm{m}}$ & $S_{25\,\mu\mathrm{m}}$ & $S_{60\,\mu\mathrm{m}}$ & $S_{100\,\mu\mathrm{m}}$ &
  D_\mathrm{C}\hphantom{01} & \llap{log}($L_\mathrm{IR}/L_\odot$\rlap{)} &  \\
  $N$~ & \colhead{Name} & \colhead{Name} & \multicolumn2c{(J2000)} &
  \colhead{(Jy)} &   \colhead{(Jy)} &   \colhead{(Jy)} &   \colhead{(Jy)} &
  \colhead{\llap{(}Mpc\rlap{)}} & \hphantom{o} & \llap{GO}A\rlap{LS} }
\colnumbers
\startdata
 001 & NGC 0034         & F00085$-$1223  & 00~11~06.6 & \llap{$-$}12~06~27 & $    0.35$ &     2.39 &    17.05 &    16.86 &  77.60 & 11.44 & * \\
 002 & NGC 0055         & NGC 0055       & 00~14~54.5 & \llap{$-$}39~11~19 & $    1.34$ &     6.25 &    77.00 &   174.09 &   3.10 & \hphantom{1}9.36 &   \\
 003 & MCG -02-01-051/2 & F00163$-$1039  & 00~18~51.4 & \llap{$-$}10~22~33 & $    0.28$ &     1.20 &     7.48 &     9.66 & 105.76 & 11.41 & * \\
 004 & NGC 0134         & F00278$-$3331  & 00~30~21.6 & \llap{$-$}33~14~38 & $    2.35$ &     2.59 &    22.65 &    58.87 &  15.90 & 10.37 &   \\
 005 & ESO 079-G003     & F00298$-$6431  & 00~32~02.0 & \llap{$-$}64~15~14 & $    0.42$ &     0.88 &     7.05 &    17.41 &  35.31 & 10.51 &   \\
 006 & NGC 0150         & F00317$-$2804  & 00~34~16.2 & \llap{$-$}27~48~10 & $    0.66$ &     1.66 &     9.66 &    17.72 &  16.20 & \hphantom{1}9.95 &   \\
 007 & NGC 0157         & F00322$-$0840  & 00~34~46.0 & \llap{$-$}08~23~53 & $    1.61$ &     2.17 &    17.93 &    42.43 &  21.92 & 10.52 &   \\
 008 & ESO 350-IG038    & F00344$-$3349  & 00~36~52.0 & \llap{$-$}33~33~19 & $    0.52$ &     2.51 &     6.88 &     5.04 &  81.20 & 11.22 & * \\
\enddata
  \tablecomments{Table~\ref{tab:iras} is published in its entirety in
    the machine-readable format.}
\end{deluxetable*}

The multipanel fig.~1 in \citet{san03} presents an atlas of Digitized
Sky Survey (DSS1) optical images for all RBGS sources with elliptical
overlays bounding the {\it IRAS} $3 \sigma$ position uncertainties.

\subsection{The 1.28\,GHz MeerKAT image atlas}

\begin{figure}[!htb]
  \centering
  \includegraphics[width=0.5\textwidth,trim = {1.5cm 5.cm 0.cm 5.cm},clip]
  {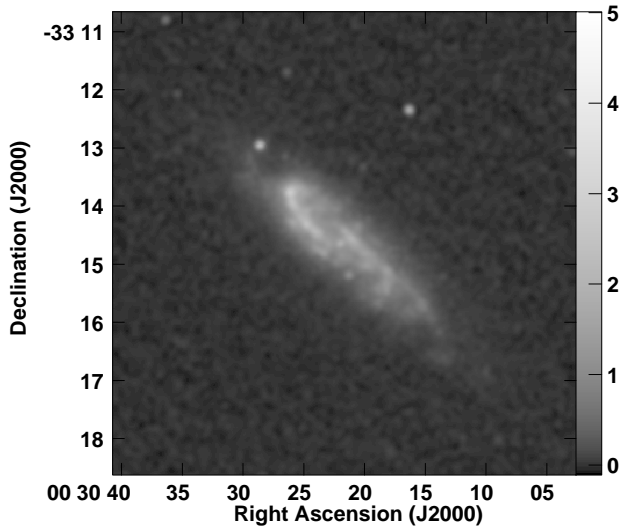}
  \caption{This MeerKAT 1.28\,GHz image of NGC 0134 is displayed over
    the intensity range $-0.1 < S_\mathrm{P}
    (\mathrm{mJy\,beam}^{-1}) < 5$ indicated by the scale bar in units
    of mJy\,beam$^{-1}$ on the right.  The complete figure set of 298
    MeerKAT images is available.
  \label{fig:ngc0134}}
\end{figure}

The cropped 1.28\,GHz Stokes {\it I} images are displayed individually
as gray-scale images in Figure Set~\ref{fig:ngc0134} with labels
indicating their source index numbers in Table~\ref{tab:iras}; e.g.,
the $N = 298$ source identified with NGC~7793 appears in Figure~2.298.
These images are also available in
  FITS format from \url
  {https://doi.org/10.48479/dnt7-6q05}

\figsetstart \figsetnum{2} \figsettitle{1.28 GHz MeerKAT images of
  southern sources in the {\it IRAS} RBGS}.\\
The full Figure Set~\ref{fig:ngc0134} is available in .pdf format from
{\url ftp://ftp.cv.nrao.edu/NRAO-staff/\allowbreak jcondon/\allowbreak RBGS.dir/MeerKAT.dir/pdf.dir/}.

\figsetend

\begin{figure}[!htb]
  \centering
  \includegraphics[trim = {1.5cm 4.5cm 0.cm 4.5cm},clip,
    width = 0.23\textwidth]{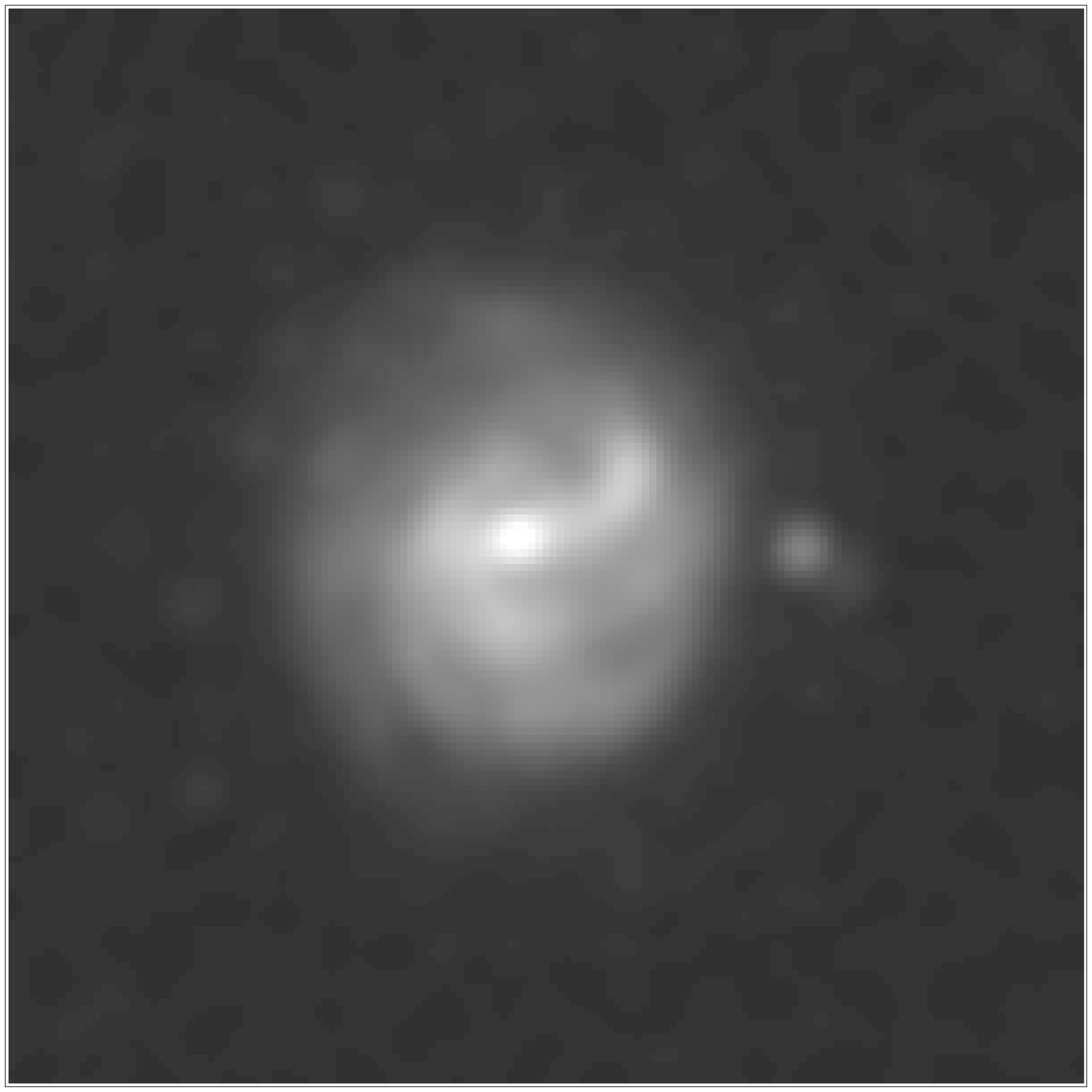}
  \includegraphics[trim = {1.5cm 4.5cm 0.cm 4.5cm},clip,
    width = 0.23\textwidth]{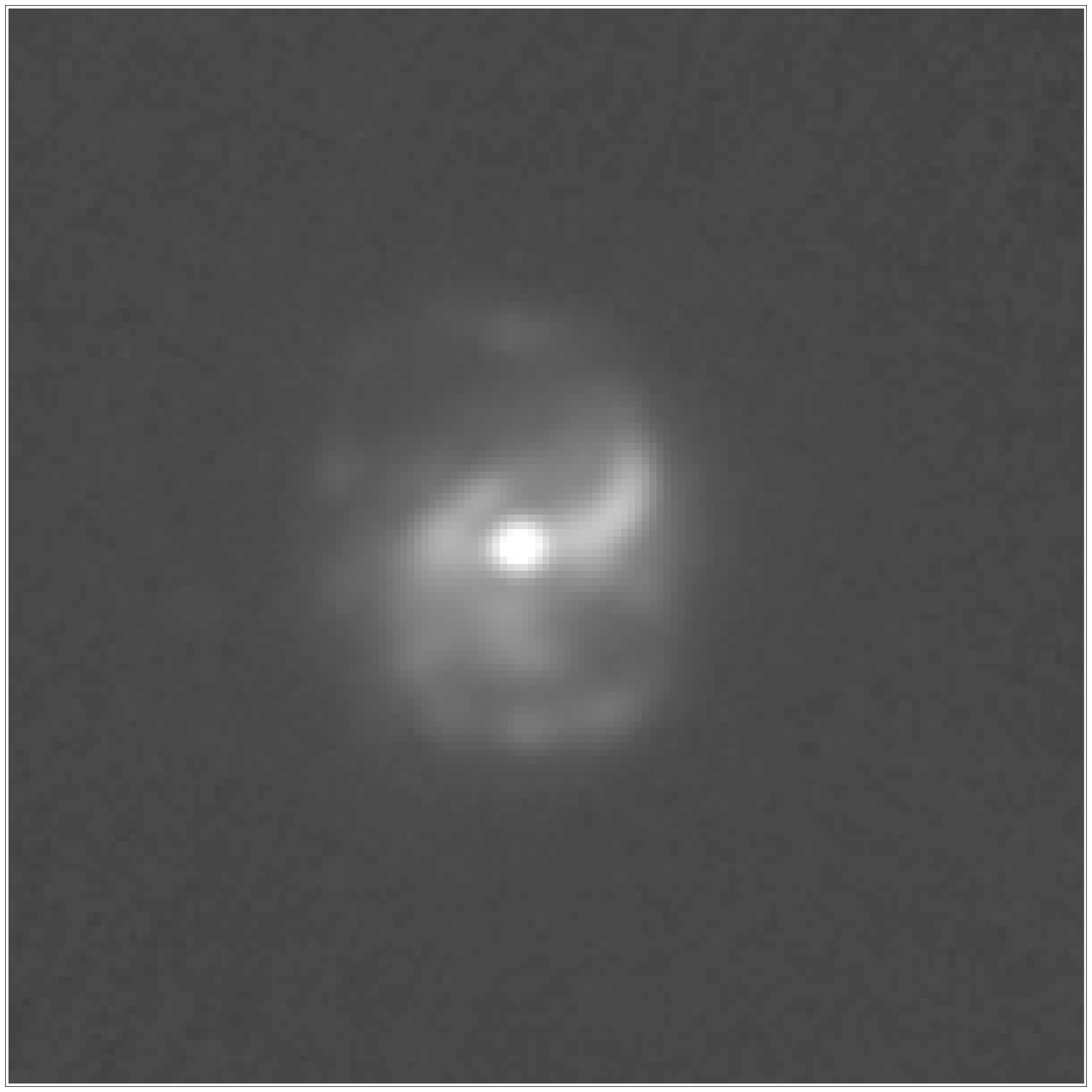}
  \caption{Comparison of $3\arcmin \times 3\arcmin$ images of NGC~6156
    from MeerKAT (left) at $\nu = 1.28\,\mathrm{GHz}$ and $7\,\farcs4$
    resolution and from {\it Herschel} (right) at $\lambda = 100
    \,\mu\mathrm{m}$ and $6\,\farcs8$ resolution \citep{chu17}.
      \label{fig:Herschel}}
\end{figure}

Figure~\ref{fig:Herschel} compares $3\arcmin \times 3\arcmin$ MeerKAT
1.28\,GHz and {\it Herschel} $\lambda = 100\,\mu\mathrm{m}$ image
cutouts centered on NGC~6156.  When observed at the same
  resolution, most RBGS galaxies have similar radio and FIR
  morphologies.  The faint radio source to the right of NGC~6156 is a
  background source with no detectable FIR counterpart, and confusing
  background AGNs are more common at radio wavelengths.

\subsection{Basic 1.28\,GHz MeerKAT source parameters}

We extracted basic parameters describing the 1.28\,GHz radio sources
identified with RBGS sources or individual galaxies in our MeerKAT
images.  The total 1.28\,GHz flux density $S$ from each {\it IRAS}
source was obtained in an aperture covering all of its radio emission
by summing over pixel brightnesses using the {\it AIPS} verb TVSTAT.
The {\it AIPS} verb MAXFIT was used to estimate the peak flux
  densities $S_\mathrm{p}$ and equinox J2000 equatorial coordinates of
  compact components embedded in bright extended emission.
  Otherwise, the {AIPS} task JMFIT was used to make Gaussian fits to
component peak flux densities $S_\mathrm{p}$, integrated flux
densities $S_\mathrm{I}$, deconvolved size parameters (FWHM major axis
$\phi_\mathrm{M}$, minor axis $\phi_\mathrm{m}$, and position angle
{\it PA} measured counterclockwise from north), and equinox J2000
equatorial coordinates.  Some of these compact components are
surrounded by faint extended disk emission, which the Gaussian
fits ignored by excluding pixels fainter than 10\% of the peak
brightness.  In most RBGS sources the disk emission is much more
  extended than the the fitted core component.  The detailed
brightness distributions of components smaller than the beam are
indeterminate, and any deconvolved component FWHM $\phi_\mathrm{M}
\lesssim 7\,\farcs5$ implies only that the second moment of the
  component brightness distribution equals that of a Gaussian with
FWHM $= \phi_\mathrm{M}$.

\begin{figure}[!htb]
  \centering
  \includegraphics[width=0.48\textwidth,trim = {4cm 9.cm 4.5cm 7.cm},clip]
  {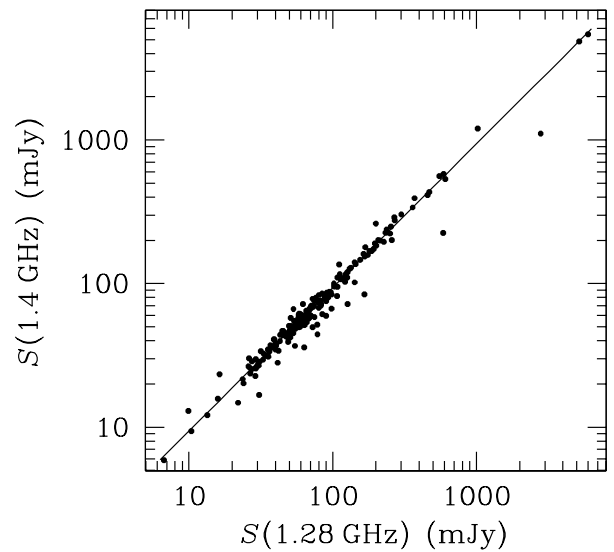}
  \caption{
    \label{fig:NVSSfluxes} The MeerKAT 1.28\,GHz flux densities and NVSS
    1.4\,GHz flux densities of RBGS galaxies north of $\delta = -40^\circ$
    cluster around the line $S(1.4\,\mathrm{GHz}) / S(1.28\,\mathrm{GHz})
    = 0.936$ expected for sources with spectral index $\alpha = -0.74$.
  }
\end{figure}

We compared our 1.28\,GHz flux densities with the 1.4\,GHz flux
densities of NVSS \citep{con98} sources north of the NVSS declination
limit $\delta > -40^\circ$ because only the NVSS images have the
surface-brightness sensitivity ($\sigma =
0.45\,\mathrm{mJy\,beam}^{-1}$ in a $\theta_\mathrm{M} =
\theta_\mathrm{m} = 45\arcsec$ beam, which corresponds to
$T_\mathrm{b} = 0.14\,\mathrm{K}$) needed to detect almost all RBGS
sources.  Most RBGS sources are dominated by optically thin
synchrotron radiation and have spectral indices $\alpha \approx -0.7$,
so we expect them to lie near the line $\langle S(1.4\,\mathrm{GHz}) /
S(1.28\,\mathrm{GHz}) \rangle \approx 0.94$ in
Figure~\ref{fig:NVSSfluxes}.

\begin{figure}[!htb]
  \centering
  \includegraphics[width=0.5\textwidth,trim = {1.5cm 5.cm 0.cm 5.5cm},clip]
  {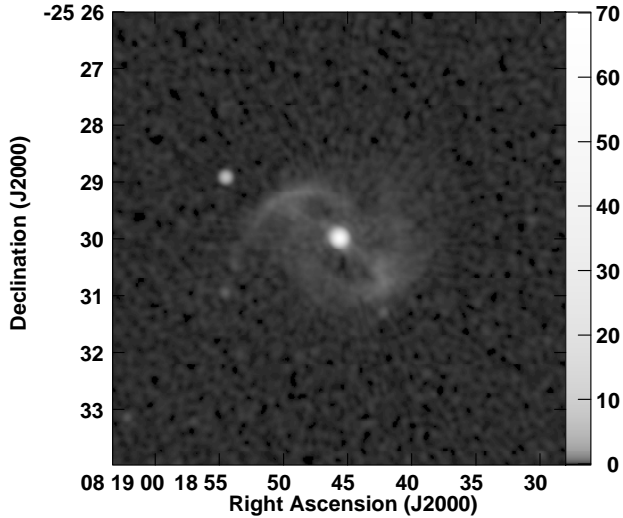}
  \caption{
    \label{fig:NGC2566} The diffuse emission surrounding
    the center of NGC~2566 visible in our 1.28\,GHz MeerKAT image is
    too faint to contribute to the NVSS cataloged flux density.  The
    scale bar on the right indicates intensity in
    $\mathrm{mJy\,beam}^{-1}$.  }
\end{figure}

The 1.4\,GHz flux densities of some NVSS sources fall significantly
below that line, usually because the NVSS catalog is incomplete for
emission fainter than $S_\mathrm{p} = 2.4\,\mathrm{mJy\,beam}^{-1}$.  For
example, the $S(1.4\,\mathrm{GHz}) = 102\,\mathrm{mJy}$ NVSS flux density
of NGC~2566 does not include the faint disk surrounding the compact
nuclear component (Figure~\ref{fig:NGC2566}), which is included in the
MeerKAT flux density $S(1.28\,\mathrm{GHz}) = 142\,\mathrm{mJy}$.

\begin{figure}[!htb]
  \centering \includegraphics[width=0.5\textwidth,trim = {4.5cm 9.cm
      4.1cm 9.5cm},clip] {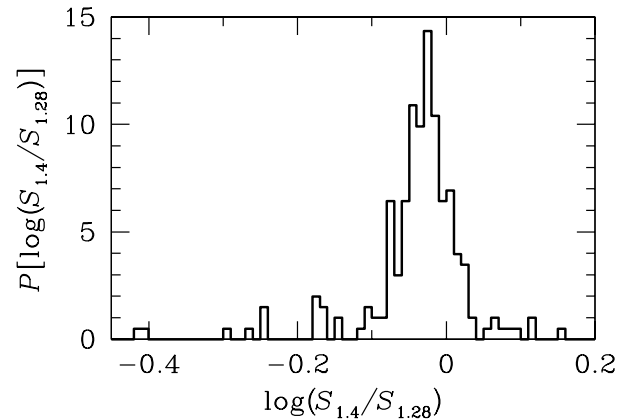}
  \caption{
    \label{fig:fluxratios} Histogram of the logarithmic flux-density ratio
    $\log[(S1.4\,\mathrm{GHz})/S(1.28\,\mathrm{GHz})]$.  The narrow
    peak indicates small intensity-proportional errors and the median
    $\langle \log[S1.4\,\mathrm{GHz}/S(1.28\,\mathrm{GHz})]\rangle =
    -0.0287$ is expected for sources with spectral index $\alpha =
    -0.74$.  }
\end{figure}

The distribution of the logarithmic flux-density ratio
$\log[S(1.4\,\mathrm{GHz})/S(1.28\,\mathrm{GHz})]$ shown by the
histogram in Figure~\ref{fig:fluxratios} has a narrow
(semi-interquartile range 0.0235) peak indicating good linearity and
rms intensity-proportional errors $\lesssim 5$\% at each frequency.
The median of the distribution is
$\langle\log[S1.4\,\mathrm{GHz}/S(1.28\,\mathrm{GHz})\rangle =
-0.0287$, which corresponds to perfect MeerKAT/NVSS flux-density scale
agreement for sources with median spectral index $\langle
\alpha(1.28\,\mathrm{GHz} / 1.4\,\mathrm{GHz}) \rangle= -0.74$.
Conversely, if the actual median spectral index is anywhere in the
range $-0.84 < \langle \alpha \rangle < -0.64$, the 1.28\,GHz MeerKAT
and 1.4\,GHz NVSS flux-density scales must agree within 1\%, which
  is even better than expected.

The VLA has imaged most RBGS galaxies north of $\delta = -45^\circ$ at
$\nu = 1.49\,\mathrm{GHz}$ at one or more resolutions between
$\theta_\mathrm{M}\approx 2\arcsec$ and $\theta_\mathrm{M} =
60\arcsec$ with rms noise in the range $0.1 <
\sigma_\mathrm{n}(\mathrm{mJy\,beam}^{-1}) < 0.2$ \citep{con90,con96}.
The flux-density agreement is good for most of the compact radio
sources, but the limited surface-brightness sensitivity often caused
the VLA to miss extended emission and yield flux densities
significantly lower than the MeerKAT flux densities.  Comparison of
deconvolved angular sizes measured in VLA images having
$\theta_\mathrm{M} \approx 2\arcsec$ resolution confirms that the
MeerKAT deconvolved angular-size estimates of sources as small as
$\phi_\mathrm{M} \sim 2\arcsec$ are reliable.

\begin{figure}[!htb]
  \centering
  \includegraphics[width=0.5\textwidth,trim = {1.5cm 5.5cm 4.cm 6.cm},clip]
  {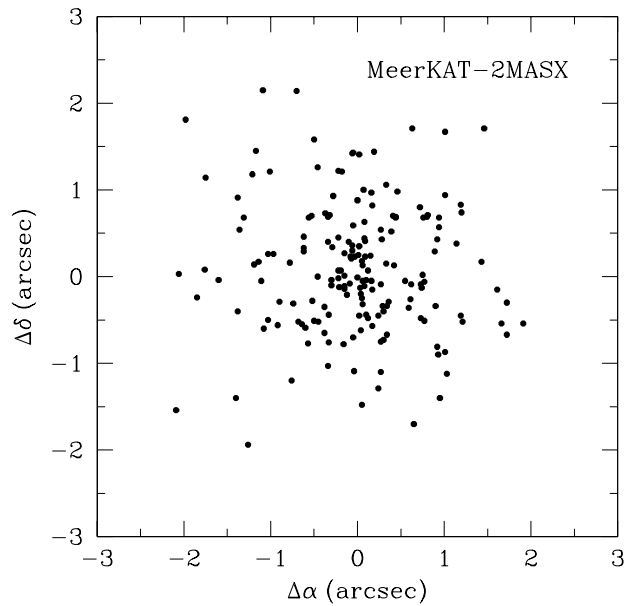}
  \caption{For galaxies with well-defined nuclei, this scatter plot
    shows the distribution of MeerKAT minus 2MASX offsets in right
    ascension $\Delta \alpha$ and declination $\Delta \delta$.
    \label{fig:2MASXoff}
  }
\end{figure}

We estimated the absolute astrometric uncertainties in our MeerKAT
1.28\,GHz image by comparing fitted positions of strong compact
sources in those images with very accurate positions in the third
realization of the International Celestial Reference Frame
\citep{cha20}.  The rms uncertainties in right ascension and
declination are $\sigma_\alpha \approx 0\,\farcs5$ and $\sigma_\delta
\approx 0\,\farcs6$, respectively.  This is larger than the
$0\,\farcs1$ we hoped for.  Possible causes include (1) large angular
separations between targets and calibration sources, (2) no on-line
interferometric refraction correction, and (3) no UT1 corrections in
the on-line delay model.

More relevant for most astronomical applications are the relative
uncertainties of our 1.28\,GHz component positions compared with their
infrared positions.  For components smaller than the MeerKAT
synthesized beam ($\phi_\mathrm{M} \lesssim 7\,\farcs5$) in galaxies
with well-defined nuclei, the distribution of MeerKAT minus accurate
($\sigma_\alpha \approx \sigma_\delta \approx 0\,\farcs1$) 2MASX
coadded positions \citep{skr06} is shown in Figure~\ref{fig:2MASXoff}.
The rms scatters in the offsets are $\sigma_{\Delta\alpha} =
1\,\farcs0$ and $\sigma_{\Delta \delta} = 1\,\farcs1$.  The mean
  offsets in right ascension and declination are $\langle \Delta
  \alpha \rangle = -0\,\farcs09 \pm 0\,\farcs07$ and $\langle \Delta
  \delta \rangle +0\,\farcs05 \pm 0\,\farcs08$, both consistent with
  zero.

Table~\ref{tab:radio} presents the 1.28\,GHz MeerKAT source parameters
of the southern RBGS sample:

{\it Column (1)}.---Our {\it IRAS} source index number $N$ from
Table~\ref{tab:iras} running from 1 through 298.

{\it Column (2)}.---The most common galaxy name(s).  If an {\it IRAS}
source contains multiple galaxies resolved by MeerKAT, the individual
galaxies are listed in additional rows with the same index number $N$.

{\it Column (3)}.---Total 1.28\,GHz MeerKAT flux density $S$ (mJy)
associated with the {\it IRAS} source.

{\it Columns (4)-(10)}.---The 1.28\,GHz component parameters
determined by fitting only the component peak or by a full Gaussian
fit.

{\it Column (4)}.---Component peak flux density in mJy per Gaussian
beam solid angle $\Omega_\mathrm{b} = \pi \theta_\mathrm{M}
\theta_\mathrm{m} / (4 \ln 2)$.

{\it Column (5)}.---Component integrated flux density (mJy) from a
Gaussian fit.

{\it Columns (6)--(8)}.---Gaussian fit deconvolved component FHWM
major axis, minor axis, and major axis position angle measured
counterclockwise from north.

{\it Columns (9)--(10)}.---Component J2000 right ascension $\alpha$
and declination $\delta$.  Positions listed with $0\,\fs01$ precision
in $\alpha$ and $0\,\farcs1$ in $\delta$ have rms uncertainties
$\sigma_\alpha = \sigma_\delta = 1\arcsec$.  Galaxies lacking clearly
defined radio nuclei have positions listed with $0\,\fs1$ precision in
$\alpha$ and $1\arcsec$ in $\delta$ and have rms uncertainties between
$1\arcsec$ and $10\arcsec$ in each coordinate.

{\it Columns (11)--(13)}.---The image restoring beam FWHM major axis
$\theta_\mathrm{M}$, minor axis $\theta_\mathrm{m}$, and major-axis
position angle {\it BPA} measured counterclockwise from north.

{\it Column (14)}.---Observing run label from Table~\ref{tab:obs}.

\startlongtable
\begin{deluxetable*}{C l R R R R R R l l R R R c}
  \centerwidetable
  \label{tab:radio}
  \tabletypesize{\small}
  \centering
  \tablecaption{The Southern RBGS Sample: 1.28\,GHz MeerKAT Data}  
  \tablehead{
    \colhead{Index} & \colhead{Common}  &  \colhead{$S$} & \colhead{$S_\mathrm{P}$} &
    \colhead{$S_\mathrm{I}$} & \colhead{$\phi_\mathrm{M}$} & \colhead{$\phi_\mathrm{m}$} &
    \colhead{{\it PA}} & \colhead{$\alpha$} & \colhead{$\delta$} &
     \colhead{$\theta_\mathrm{M}$} & \colhead{$\theta_\mathrm{m}$} &
    \colhead{{\it \llap{B}P\rlap{A}}}  & \\
    $N$ & \colhead{Name} &     \colhead{(mJy)} &
    \colhead{\llap{(}mJy/$\Omega_\mathrm{b}$\rlap{)}} & \colhead{(mJy)} &
    \colhead{($\arcsec$)} & \colhead{($\arcsec$)} & \colhead{($^\circ$)} &  \multicolumn2c{(J2000)}  &
    \colhead{($\arcsec$)} & \colhead{($\arcsec$)} & \colhead{($^\circ$)} &
    \colhead{\llap{R}u\rlap{n}}
  }
  \decimals
  \colnumbers
\startdata
 001 & \llap{NG}C 0034         & 98.0    & 58.5    & 61.6    & 1.9   & 1.7   & 4     & 00 11 06.57 & \llap{$-$}12 06 27.2 & 7.68 & 7.67 & -1   & H    \\
 002 & \llap{NG}C 0055         & 587.    & 7.74    & 16.3    & 8.3   & 7.4   & -79   & 00 14 57.54 & \llap{$-$}39 12 25.8 & 7.48 & 7.22 & 89   & K    \\
 003 & \llap{MC}G -02-01-051\rlap{/2} & 45.2 &  &  &  &  &  &  &  & 7.68 & 7.67 & -1   & H    \\
 003 & \llap{MC}G -02-01-051   &         & 30.7    & 40.0    & 4.8   & 3.6   & -54   & 00 18 50.88 & \llap{$-$}10 22 36.6 & 7.68 & 7.67 & -1   & H    \\
 003 & \llap{MC}G -02-01-052   &         & 2.8     &         &       &       &       & 00 18 49.99 & \llap{$-$}10 21 29.9 & 7.68 & 7.67 & -1   & H    \\
 004 & \llap{NG}C 0134         & 227.    & 1.8     &         &       &       &       & 00 30 21.96 & \llap{$-$}33 14 38.0 & 7.57 & 7.37 & -28  & K    \\
 005 & \llap{ES}O 079-G003     & 61.7    & 14.8    & 23.5    & 7.9   & 3.2   & -50   & 00 32 02.36 & \llap{$-$}64 15 10.9 & 7.61 & 7.34 & -1   & C    \\
 006 & \llap{NG}C 0150         & 58.2    & 17.4    & 21.0    & 3.6   & 3.0   & 80    & 00 34 15.46 & \llap{$-$}27 48 12.5 & 7.59 & 7.25 & 0    & H    \\
\enddata
\tablecomments{Table~\ref{tab:radio} is published in its entirety in
  the machine-readable format.}
\end{deluxetable*}

\subsection{Notes on individual sources}

001:~{\it NGC~0034}.---Marginally detected extremely faint ($\langle
S_\mathrm{P} \rangle \sim 15 \,\mu\mathrm{Jy\,beam}^{-1}$)
double-lobed emission with LAS $\approx 13'$ in PA $\approx 48^\circ$
is centered on the unresolved radio core.  Late-stage merger galaxy
\citep{fer10}.

002:~{\it NGC~0055}.---The radio emission from this edge-on galaxy
has  LAS = $28'$ in PA = $-73^\circ$.  The brightest radio component
listed in Table~\ref{tab:radio} at $\alpha =
00^\mathrm{h}\,14^\mathrm{m}\,57\,\fs54$, $\delta =
-39^\circ\,12\arcmin\,25\,\farcs8$ coincides with the brightest nearby
{\it WISE} source ($W1 = 9.459$, $W2 = 9.319$, $W3 = 5.312$) at
$\alpha = 00^\mathrm{h}\,14^\mathrm{m}\,57\,\fs60$, $\delta =
-39^\circ\,12\arcmin\, 25\,\farcs5$.  This may be the actual obscured
nucleus of NGC~0055, despite being significantly offset from the 2MASX
position $\alpha = 00^\mathrm{h}\,14^\mathrm{m}\,53\,\fs60$, $\delta =
-39^\circ\,11\arcmin\,47\,\farcs9$ preferred by NED.

004:~{\it NGC~0134}.---The radio source LAS = $7'$ in PA = 48$^\circ$.

007:~{\it NGC~0157}.---No visible radio nucleus.

008:~{\it ESO 350-IG038}.---Compact group of galaxies \citep{sch06}
only marginally resolved by MeerKAT. Warm $\alpha(25\,\mu\mathrm{m},
60\,\mu\mathrm{m}) = -1.15$ suggests a significant AGN
contribution to the {\it IRAS} source \citep{deg85}.

010:~{\it NGC~0232}.---The companion galaxy NGC~0230 is a $S =
9.5\,\mathrm{mJy}$ radio source with FWHM size $\phi_\mathrm{M} =
16\,\farcs0$ and $\phi_\mathrm{m} = 4\,\farcs9$ in PA = $43^\circ$
centered on $\alpha = 00^\mathrm{h}\,42^\mathrm{m}\,27\,\fs14$,
$\delta = -23^\circ\,37\arcmin\, 43\,\farcs44$.  The companion galaxy
NGC~0232E is an unresolved radio source with $S = 39.2$\,mJy at
$\alpha = 00^\mathrm{h}\, 42^\mathrm{m}\, 52\,\fs83$, $\delta =
-23^\circ\,32\arcmin \, 27\,\farcs4$.

011:~{\it NGC~0247}.---Extremely diffuse and patchy galaxy, marginal
radio detection, very uncertain total flux density and position.

012:~{\it NGC~0253}.---The radio source LAS = $19'$ in PA = $49^\circ$.

014:~{\it NGC~0300}.---Very faint patchy radio source, no visible
nucleus, very uncertain total flux density and position.  LAS $\sim
17'$.

021:~{\it NGC~0625}.---The radio continuum from this dwarf starburst
galaxy is primarily free-free emission \citep{can04}.  The MeerKAT
position is for the brightest of three \ion{H}{2} regions, about
$2^\mathrm{s}$ east of the galaxy centroid.

026:~{\it NGC~0835}.---The nearby galaxy NGC~0833 is a 4.3\,mJy
compact radio source at $\alpha =
02^\mathrm{h}\,09^\mathrm{m}\,20\,\fs84$, $\delta =
-10^\circ\,07\arcmin\,59\,\farcs6$ and may contribute to the {\it
  IRAS} source.

031:~{\it NGC~0922}.---Collisional ring galaxy \citep{ela18}, cometary
morphology with bright eastern rim.  The nucleus is offset to the
northeast.

035:~{\it NGC~1068}.---Warm $\alpha(25\,\mu\mathrm{m},
60\,\mu\mathrm{m}) = -0.92$ and low $q = 1.72$ indicate a dominant AGN
contribution \citep{deg85} to the radio flux of this starburst galaxy.

038:~{\it NGC~1087}.---The compact $S = 12$\,mJy source at
$\alpha = 02^\mathrm{h}\,46^\mathrm{m}\,29\,\fs28$,
$\delta = -00^\circ\,29\arcmin\,52\,\farcs0$ superimposed on the
eastern edge of NGC~1087 appears to be an unrelated background source.

039:~{\it NGC~1097}.---Circumnuclear radio ring \citep{hum87}.
The MeerKAT central brightness $S_\mathrm{p} = 15\,
\mathrm{mJy\,beam}^{-1}$ is less than the $S_\mathrm{p} = 21\,
\mathrm{mJy\,beam}^{-1}$ of the surrounding ring, so the Gaussian
fitted angular size is an overestimate.

043:~{\it NGC~1232}.---The center of the radio image is empty, with no
detectable nucleus.

045:~{\it NGC~1313}.---No clear radio nucleus, very uncertain position.

046:~{\it NGC~1309}.---No clear radio nucleus, very uncertain position.
The compact $S_\mathrm{p} = 3.4\,\mathrm{mJy\,beam}^{-1}$ source at
$\alpha = 03^\mathrm{h}\,22^\mathrm{m}\,06\,\fs3$,
$\delta = -15^\circ\,23\arcmin\,18\arcsec$ behind the north edge
of NGC~1309 is not included in the flux density.

048:~{\it IC~1953}.---Compact radio core surrounded by a faint
circular halo.

050:~{\it NGC~1377}.---Exceptionally faint radio source ($q = 4.33$).
The marginally resolved radio continuum extends parallel to the
molecular jet \citep{aal20}.

051:~{\it NGC~1386}.---Faint radio disk or jets lie nearly
perpendicular to the 2MASX stellar disk.

058:~{\it NGC~1511}.---Fitted position may not be the actual nucleus.
Bright eastern radio arc.

059:~{\it NGC~1532}.---Thin edge-on disk with perpendicular radio
plumes suggesting outflows.  Interacting with NGC~1531, an $S =
2.4$\,mJy source at $\alpha = 04^\mathrm{h}\,59^\mathrm{m}\,
59\,\fs26$, $\delta = -32^\circ\,51\arcmin\,04\,\farcs2$ with FWHM
size $16\,\farcs9 \times 14\,\farcs5$ in PA $= -57^\circ$

061:~{\it NGC~1546}.---The optically faint compact $S =
35\,\mathrm{mJy}$ source at $\alpha =
04^\mathrm{h}\,14^\mathrm{m}\,35\,\fs31$, $\delta =
-56^\circ\,03\arcmin\,44\,\farcs2$ is not included in the NGC~5146
flux density.

062:~{\it IC~2056}.---Radio ring with central hole on the position of
the 2MASX stellar nucleus.

063:~{\it NGC~1559}.---Bright, patchy radio emission from the spiral
arms, no clearcut nucleus.

064:~{\it NGC~1566}.---Bright, compact Seyfert 1 nucleus and clear
spiral arms.

065:~{\it ESO~550-IG025}.---Merging pair of galaxies.

067:~{\it NGC~1614}.---Late-stage merger.  A faint ($S_\mathrm{p} <
0.1\,\mathrm{mJy\,beam}^{-1}$) curved tail about $2\arcmin$ long and
$\leq 7^\mathrm{s}$ east of NGC~1614 contributes $S \approx
4\,\mathrm{mJy}$ to the flux density, and the base of the straight
tail in PA $\approx -150^\circ$ is also visible.

068:~{\it ESO~485-G003}.---The companion edge-on disk galaxy
ESO~485-G004 with flux density $S = 3.3\,\mathrm{mJy}$ at $\alpha =
04^\mathrm{h}\,39^\mathrm{m}\,06\,\fs39$, $\delta =
-24^\circ\,11\arcmin\,03\,\farcs3$ probably contributes to the {\it
  IRAS} source.

071:~{\it ESO~203-IG001}.---Completely unresolved radio source, FWHM
$< 1\arcsec$.

076:~{\it NGC~1792}.---The FR\,II source at $\alpha =
04^\mathrm{h}\,20^\mathrm{m}\,15\,\fs16$, $\delta =
-54^\circ\,53\arcmin\,46\,\farcs3$ has no bright optical counterpart
and is probably an unrelated background galaxy.

079:~{\it NGC~1832}.---The compact $S = 4.5\,\mathrm{mJy}$ source
at $\alpha =  05^\mathrm{h}\,12^\mathrm{m}\,03\,\fs50$,
$\delta = -15^\circ\,41\arcmin\,32\,\farcs7$ appears to be
an unrelated background source and was not included in the
flux density of NGC~1832.

082:~{\it NGC~1964}.---No distinct radio nucleus is visible in the
bright central region.

083:~{\it NGC~2076}.---No radio nucleus.

084:~{\it NGC~2139}.---No distinct radio nucleus.

087:~{\it NGC~2207/IC~2163}.---Table 1 in \citet{san03} lists the
identification of IRAS F06142$-$2121 as IC~2163 only, but at a
position closer to that of NGC~2207.  Our 1.28\,GHz MeerKAT image
suggests that F06142$-$2121 is a blend of both galaxies in this
merging pair.  Our total 1.28\,GHz flux density $S = 371
\,\mathrm{mJy}$ includes both NGC~2207 and IC~2163 but excludes the
apparently unrelated $S = 8\,\mathrm{mJy}$ compact source overlapping
the western edge of NGC~2207 at $\alpha =
06^\mathrm{h}\,16^\mathrm{m}\,15\,\fs90$, $\delta =
-21^\circ\,22\arcmin\,03\,\farcs1$.

088:~{\it UGCA~127}.---There are several bright spots near the center
of UGCA~127, but none is clearly the nucleus.

090:~{\it NGC~2221}.---Interacting pair with NGC~2222, $S =
15.8\,\mathrm{mJy}$ at $\alpha =
06^\mathrm{h}\,20^\mathrm{m}\,16\,\fs99$, $\delta =
-57^\circ\,32\arcmin\, 04\,\farcs4$.

091:~{\it ESO~005-G004}.---The MeerKAT nuclear position
$\alpha = 06^\mathrm{h}\,05^\mathrm{m}\,38\,\fs2$,
$\delta =  -86^\circ\,37\arcmin\,53\arcsec$ formally appears to be 
offset from the 2MASX coadd position
$\alpha = 06^\mathrm{h}\,05^\mathrm{m}\,41\fs36$,
$\delta = -86^\circ\,37\arcmin\,54\,\farcs2$. However, it is
very near the south celestial pole and is probably inaccurate.

092:~{\it ESO~255-IG007}.---Tight galaxy triplet.

093:~{\it ESO~557-G002}.---Pair with ESO~557-G001, $S =
11.3\,\mathrm{mJy}$ at $\alpha =
06^\mathrm{h}\,31^\mathrm{m}\,45\,\fs73$, $\delta =
-17^\circ\,38\arcmin\,47\,\farcs5$.

097:~{\it AM~0702$-$601}.---Pair of compact galaxies.

098:~{\it ESO~491-G020/021}.---Interacting pair of galaxies.

099:~{\it ESO~492-G002}.---Pair with ESO~492-G003, $S =
4.6\,\mathrm{mJy}$ at $\alpha =
07^\mathrm{h}\,11^\mathrm{m}\,45\,\fs01$, $\delta =
-26^\circ\,39\arcmin\,30\,\farcs3$.

101:~{\it ESO~428-G023}.---The $S_\mathrm{p} =
3.2\,\mathrm{mJy\,beam}^{-1}$ source at $\alpha =
07^\mathrm{h}\,22^\mathrm{m}\,06\,\fs39$, $\delta =
-29^\circ\,14\arcmin\,04\,\farcs2$ is optically faint and probably an
unrelated background galaxy.

105:~{\it NGC~2442}.---Large open spiral with polarized continuum
wisps extending east from the stellar image \citep{har04}, relatively
radio loud ($q = 1.96$).

108:~{\it NGC~2525}.---The optically bright stellar bar is not visible
in the MeerKAT image.

112:~{\it ESO~432-IG006}.---Tidally interacting pair of galaxies.

115:~{\it ESO~060-IG016}.---Close pair of galaxies.

117:~{\it ESO~564-G011}.---Our flux density $S = 78.7\,\mathrm{mJy}$
does not include the $S = 5.3\,\mathrm{mJy}$ companion galaxy
ESO~564-G010 at $\alpha = 09^\mathrm{h}\,02^\mathrm{m}\,45\,\fs20$,
$\delta = -20^\circ\,42\arcmin\,49\,\farcs7$, which may contribute to
the {\it IRAS} source.

120.~{\it IRAS~F09111$-$1007}.---The {\it IRAS} source is a blend
of the two galaxies {\it IRAS}~F09111$-$1007W and
{\it IRAS}~F09111$-$1007E.

123.~{\it NGC~2992}.---Interacting pair with NGC~2993.

124.~{\it NGC~2993}.---Interacting pair with NGC~2992.

125.~{\it NGC~3059}.---The $S = 117\,\mathrm{mJy}$ flux density
excludes the compact 6.8\,mJy source at $\alpha =
09^\mathrm{h}\,50^\mathrm{m}\,18\,\fs77$, $\delta =
-73^\circ\,55\arcmin\,30\,\farcs2$.  Confusing radio source near the
nucleus?

126.~{\it IC~2522}.---The $S = 56.7\,\mathrm{mJy}$ flux density
excludes the 25\,mJy \ companion galaxy IC 2523 at $\alpha =
09^\mathrm{h}\,55^\mathrm{m}\,09\,\fs54$, $\delta =
-33^\circ\,12\arcmin\,37\,\farcs0$.

127.~{\it NGC~3095}.---NGC~3095 is a member of the NGC~3100 ($S =
614\,\mathrm{mJy}$ at $\alpha =
10^\mathrm{h}\,00^\mathrm{m}\,40\,\fs84$, $\delta =
-31^\circ\,39\arcmin\,51\,\farcs6$) group.

128.~{\it NGC~3110}.---The $S = 131\,\mathrm{mJy}$ flux density
includes neither the 6.4\,mJy companion galaxy MCG~-01-26-013 at
$\alpha = 10^\mathrm{h}\,03^\mathrm{m}\,57\,\fs05$, $\delta =
-06^\circ\,29\arcmin\,47\,\farcs6$ nor the small nearby source at
$\alpha = 10^\mathrm{h}\,04^\mathrm{m}\,00\,\fs05$, $\delta =
-06^\circ\,28\arcmin\,17\,\farcs5$.

129.~{\it ESO~374-IG032}.---This galaxy pair is the correct
identification of the {\it IRAS} source F10038$-$3338, which had been
incorrectly identified with IC~2545.

130.~{\it NGC~3125}.---Blue compact dwarf galaxy.  Two bright areas are
resolved in both the 2MASX {\it JHK}$_\mathrm{s}$ stellar image and in
our 1.28\,GHz MeerKAT image.

136.~{\it NGC~3263}.---Long eastern radio tail.  Interacting pair with
the radio-quiet galaxy NGC~3262 at $\alpha =
10^\mathrm{h}\,29^\mathrm{m}\,06\,\fs23$, $\delta =
-44^\circ\,09\arcmin\,34\,\farcs8$, member of the NGC~3256 group.

137.~{\it NGC~3278}.---Cometary radio morphology brightest on the
northwestern edge.  No visible radio nucleus.

138.~{\it NGC~3281}.---Radio jets extend perpendicular to
the disk of this radio-loud ($q = 1.99$) Seyfert 2 galaxy

140.~{\it ESO~264-G057}.---Brightest in group including ESO~264-G058
($S = 6.0\,\mathrm{mJy}$ at $\alpha =
10^\mathrm{h}\,59^\mathrm{m}\,06\,\fs9$, $\delta =
-43^\circ\,22\arcmin\,34\arcsec$) and WISEA J105911.26-432826.7 ($S =
3.8\,\mathrm{mJy}$ at $\alpha =
10^\mathrm{h}\,59^\mathrm{m}\,11\,\fs4$, $\delta =
-43^\circ\,28\arcmin\,29\arcsec$).

143.~{\it NGC~3511}.---Pair with NGC~3513 ($S = 21.1\,\mathrm{mJy}$ at
$\alpha = 11^\mathrm{h}\,03^\mathrm{m}\,46\,\fs3$, $\delta =
-23^\circ\,14\arcmin\,45\arcsec$).

144.~{\it ESO~265-G007}.---No visible radio nucleus.

146.~{\it NGC~3568}.---Large (LAS $= 19\arcmin$) tailed $S =
1.01\,\mathrm{Jy}$ radio galaxy NGC~3557 superimposed.

149.~{\it NGC~3621}.---Patchy irregular galaxy with no recognizable
radio nucleus.

150.~{\it CGCG~011-076}.---The $S = 39.9\,\mathrm{mJy}$ flux density
does not include the possibly interacting companion galaxy LCRS
B111835.0$-$024314 ($S = 1.2\,\mathrm{mJy}$ at $\alpha =
11^\mathrm{h}\,21^\mathrm{m}\,08\,\fs29$, $\delta =
-02^\circ\,59\arcmin\,39\,\farcs4$.

153.~{\it NGC~3717}.---Pair with IC~2913 ($S = 9.8\,\mathrm{mJy}$ at
$\alpha = 11^\mathrm{h}\,31^\mathrm{m}\,51\,\fs2$, $\delta =
-30^\circ\,24\arcmin\,41\arcsec$).

161.~{\it NGC~4027}.---No clearcut radio nucleus.

162.~{\it NGC~4030}.---No visible radio nucleus.

163.~{\it NGC~4038/9}.---NGC~4039 has no clearcut radio nucleus.

164.~{\it ESO~440-IG058}.---Interacting pair of galaxies.

165.~{\it ESO~267-G030}.---Pair with ESO~267-G029 ($S =
30.9\,\mathrm{mJy}$ at $\alpha =
12^\mathrm{h}\,13^\mathrm{m}\,52\,\fs23$, $\delta =
-47^\circ\,16\arcmin\,25\,\farcs6$).

171.~{\it NGC~4418}.---Seyfert 2 galaxy.  Radio source FWHM $<
1\arcsec$ and unusually high $q = 3.12$ suggest this radio source is
AGN dominated.

172.~{\it NGC~4433}.---Pair with NGC~4428 ($S = 52.0 \,\mathrm{mJy}$
at $\alpha = 12^\mathrm{h}\,27^\mathrm{m}\,24\,\fs09$,
$\delta = -08^\circ\,10\arcmin\,52\,\farcs2$).

174.~{\it IC~3639}.---Seyfert 2 galaxy with companion galaxies
ESO~381-G006 ($S = 0.8\,\mathrm{mJy}$ at
$\alpha = 12^\mathrm{h}\,40^\mathrm{m}\,40\,\fs79$,
$\delta = -36^\circ\,44\arcmin\,21\,\farcs5$) and
ESO~381-G009 ($S = 15.7\,\mathrm{mJy}$ at
$\alpha = 12^\mathrm{h}\,40^\mathrm{m}\,58\,\fs34$,
$\delta = -36^\circ\,43\arcmin\,54\,\farcs0$).

175.~{\it NGC~4666}.---Edge-on spiral galaxy with radio emission
extending well above and below the disk.

176.~{\it NGC~4691}.---There are two brightness peaks in the MeerKAT
radio image, which we labeled NGC~4691A and NGC~4691B.

177.~{\it NGC~4699}.---Compact radio core surrounded by a very
faint ($\sim 0.1\,\mathrm{mJy\,beam}^{-1}$) halo with FWHM $=3\arcmin$.

178.~{\it NGC~4781}.---No detectable radio nucleus.

183.~{\it MCG~-02-33-098/9}.---IRAS F12596$-$1529 is a blend of
MCG~-02-33-098W and MCG~-02-33-098.  MCG~-02-33-099 is radio quiet

193.~{\it NGC~5054}.---Interacting pair with MCG~-03-34-040
($S_\mathrm{p} = 0.2 \,\mathrm{mJy\,beam}^{-1}$ at
$\alpha = 13^\mathrm{h}\,16^\mathrm{m}\,56\,\fs2$,
$\delta = -16^\circ\,35\arcmin\,31\arcsec$.

194.~{\it NGC~5068}.---Diffuse source with no visible radio nucleus.

196.~{\it NGC~5078}.---Edge-on disk with symmetric perpendicular radio
jets or winds originating in the nucleus.

197.~{\it MCG~-03-34-064}.---AGN emission dominates the radio emission
from this unresolved radio-loud ($q = 1.47$) warm
[$\alpha(25\,\mu\mathrm{m},60\,\mu\mathrm{m}) = -0.84$] Seyfert 1.8
galaxy.

198.~{\it NGC~5128}.---Cen A radio galaxy.

203.~{\it NGC~5236}.---M83.

204.~{\it NGC~5247}.---No clearcut radio nucleus.

206.~{\it ESO~221-IG008}.---Marginally resolved pair of galaxies.

208.~{\it NGC~5427}.---Pair with NGC~5426 ($S = 45\,\mathrm{mJy}$,
$3'$ south, no visible radio nucleus).

209.~{\it NGC~5483}.---Central bulge but no visible nucleus.  Flux density
excludes three compact sources on the northwest side.

211.~{\it NGC~5506}.---The strong ($q = 1.48$) compact radio source in
this warm [$\alpha(25\,\mu\mathrm{m},60\,\mu\mathrm{m}) = -0.80$]
Seyfert 1.9 nucleus is probably dominated by AGN emission.

213.~{\it NGC~5595}.---Pair with NGC~5597.

214.~{\it NGC~5597}.---Pair with NGC~5595.

215.~{\it IC~4444}.---IC~4444 = IC~4441.

218.~{\it NGC~5713}.---Interacting pair with NGC~5719.  Indistinct
offset radio nucleus.

220.~{\it NGC~5719}.---Interacting pair with NGC~5713.

222.~{\it NGC~5734}.---Pair with NGC~5734S ($S = 58.3\,\mathrm{mJy}$
at $\alpha = 14^\mathrm{h}\,45^\mathrm{m}\,11\,\fs15$,
$\delta = -20^\circ\,54\arcmin\,49\,\farcs1$).

226.~{\it IC~4518A/B}.---Interacting galaxy pair, possible AGN
contribution to the relatively loud ($q = 1.85$) radio source.

228.~{\it NGC~5792}.---The $S = 78.1\,\mathrm{mJy}$ flux density
excludes the $S = 5.5\,\mathrm{mJy}$ compact source at
$\alpha = 14^\mathrm{h}\,58^\mathrm{m}\,51\,\fs49$,
$\delta = -42^\circ\,00\arcmin\,34\,\farcs8$.

229.~{\it NGC~5793}.---Seyfert 2, strong ($q = 0.96$) flat-spectrum
radio source OQ~194 dominated by AGN emission, near NGC~5796.

233.~{\it NGC~5915}.---Triplet with NGC~5916 and NGC~5916A.

234.~{\it ESO~099-G004}.---Marginally resolved interacting pair of
galaxies?

240.~{\it ESO~452-G005}.---Pair with ESO~452-G007 ($S =
19.3\,\mathrm{mJy}$ at $\alpha =
16^\mathrm{h}\,32^\mathrm{m}\,03\,\fs28$, $\delta =
-28^\circ\,05\arcmin\,36\,\farcs0$).  There is no published
{\it IRAS} $\lambda = 100\,\mu\mathrm{m}$ flux density for
this source.

243.~{\it IRAS~F16399$-$0937}.---Marginally resolved pair of galaxies.

244.~{\it ESO~453-G005}.---Pair with
WISEA~J164729.33\allowbreak$-$291906.6 ($S = 14.6\,\mathrm{mJy}$ at
$\alpha = 16^\mathrm{h}\,47^\mathrm{m}\,29\,\fs37$, $\delta =
-29^\circ\,19\arcmin\,05\,\farcs4$).

245.~{\it NGC~6215}.---Pair with NGC~6221.

246.~{\it NGC~6221}.---Pair with NGC~6215.

248.~{\it NGC~6300}.---Seyfert 2.

252.~{\it IC~4662}.---Irr galaxy with bright \ion{H}{2} regions and no
visible radio nucleus.

253.~{\it IRAS~17578$-$0400}.---In group of three galaxies, all of
which may contribute to the {\it IRAS} source.

254.~{\it IC~4687/6}.---IC~4689 ($S = 24.7\,\mathrm{mJy}$ at $\alpha =
18^\mathrm{h}\,13^\mathrm{m}\,40\,\fs22$, $\delta =
-57^\circ\,44\arcmin\,53\,\farcs1$) might contribute slightly to the
{\it IRAS} source.

256.~{\it IRAS~F18293$-$3413}.---The $S_\mathrm{p} =
13.5\,\mathrm{mJy\,beam}^{-1}$ unresolved radio source at $\alpha =
18^\mathrm{h}\,32^\mathrm{m}\,44\,\fs86$, $\delta =
-34^\circ\,13\arcmin\,44\,\farcs3$ appears to be an unrelated
optically faint background galaxy.

258.~{\it NGC~6744}.---This very extended (LAS $\approx 15'$) galaxy
has a very faint ($S = 0.26\,\mathrm{mJy}$) unresolved radio nucleus.

259.~{\it NGC~6753}.---The $S = 148\,\mathrm{mJy}$ flux density does
not include the $S = 3.5\,\mathrm{mJy}$ compact  source
at $\alpha = 19^\mathrm{h}\,11^\mathrm{m}\,20\,\fs0$, $\delta =
-57^\circ\,02\arcmin\,37\,\farcs6$.

260.~{\it ESO~593-IG008}.---Merging pair of galaxies.

265.~{\it NGC~6822}.---Our 1.28\,GHz flux density and position of this
large, patchy, low-brightness irregular galaxy are very uncertain.
See \citet{can06} for a detailed study of the FIR and radio emission
from NGC~6822.

267.~{\it ESO~339-G011}.---Radio-loud ($q = 1.73$) Seyfert 2 galaxy.

272.~{\it IC~5063.}---Radio-loud ($q = 0.67$) warm
[$\alpha(25\,\mu\mathrm{m}, 60\,\mu\mathrm{m}) = -0.46$] Seyfert 2 AGN
core with radio plume.  Serendipitous background ($V_\mathrm{h} =
18052\,\mathrm{km\,s}^{-1}$) giant radio galaxy at $\alpha =
20^\mathrm{h}\,51^\mathrm{m}\,39\,\fs8$, $\delta =
-57^\circ\,04\arcmin\,34\arcsec$ (L.~Marchetti et al. 2021, in prep).

276.~{\it NGC~7083}.---No clearcut radio nucleus.

277.~{\it NGC~7090}.---Asymmetric edge-on spiral disk, no clearcut
radio nucleus.

278.~{\it ESO~343-IG013}.---Interacting pair of galaxies.

281.~{\it NGC~7205}.---No visible radio nucleus.

286.~{\it ESO~239-IG002}.---Unresolved merging pair of galaxies.

290.~{\it ESO~18-IG002}.---Unresolved merging pair of galaxies.

292.~{\it NGC~7582}.---Seyfert 2.  Background radio-loud BL Lac object
MCG 07-47-031 at $\alpha = 23^\mathrm{h}\,19^\mathrm{m}\,05\,\fs93$,
$\delta = -42^\circ\,06\arcmin\,48\,\farcs3$.

293.~{\it NGC~7592}.---Interacting triple system.

294.~{\it NGC~7590}.---Seyfert 2.

296.~{\it ESO~077-IG014}.---Interacting pair of galaxies.

298.~{\it NGC~7793}.---Faint extended emission with numerous
\ion{H}{2} regions superimposed, no clearcut radio nucleus.

\section{Summary}
\label{sec:summary}

The MeerKAT Atlas comprises 1.28\,GHz images of the 298 RBGS external
galaxies with $\vert b \vert > 5^\circ$ in the southern hemisphere
that are stronger than $S = 5.24\,\mathrm{Jy}$ at $\lambda =
60\,\mu$m.  These images have $\theta \approx 7\,\farcs5$ FWHM
resolution, or $\approx 1.8$\,kpc at the median angular-size distance
$\langle D_\mathrm{A} \rangle \approx 50\,\mathrm{Mpc}$ of RBGS
galaxies.  Each galaxy was observed with only 5 $\times$ 3\,min
snapshots, but the large number of MeerKAT antennas (64) gave good
$(u,v)$-plane coverage and enabled accurate imaging of complex
sources.  The central portions of these images in FITS format can be
downloaded from \url{https://doi.org/10.48479/dnt7-6q05}.

The typical rms image fluctuation $\sigma \approx
20\,\mu\mathrm{Jy\,beam}^{-1}$ is a combination of $\sigma_\mathrm{n}
\approx 15\,\mu\mathrm{Jy\,beam}^{-1}$ thermal noise,
  $\sigma_\mathrm{c} \approx 2\,\mu\mathrm{Jy\,beam}^{-1}$ rms
  confusion from numerous faint sources, and residual sidelobes from
strong sources in the $\Theta_{1/2} \approx 68\arcmin$ FWHM primary
beam.  The corresponding surface-brightness noise $\sigma \approx
0.26\,\mathrm{K}$ at 1.28\,GHz is low enough to reveal most disk
emission from SFGs with about the same angular resolution as {\it
  Herschel} at $\lambda = 100\,\mu$m or {\it WISE} at mid-infrared
wavelengths.  Thus each MeerKAT image is sufficient to completely
image a nearby galaxy in detail, and many images reveal low-brightness
features not previously seen. In contrast, early VLA atlases
\citep{con90,con96} required two or more images per galaxy, a
high-resolution image to isolate and resolve compact components and a
low-resolution image to detect low-brightness extended emission.

The Atlas images are based on $5 \times 3\,\mathrm{min}$ snapshots, so
full tracks on individual sources would bring the rms image
fluctuations much closer to the confusion limit $\sigma \gtrsim
\sigma_\mathrm{c} \approx 2\,\mu\mathrm{Jy\,beam}^{-1}$.  $S$-band
receivers covering 1.75 to 3\,GHz currently being added to MeerKAT
will lengthen the spectral baseline to allow more accurate
measurements of continuum spectral indices and also bring higher angular
resolution.

\begin{acknowledgments}
The MeerKAT telescope is operated by the South African Radio Astronomy
Observatory, which is a facility of the National Research Foundation,
an agency of the Department of Science and Innovation.  The National
Radio Astronomy Observatory is a facility of the National Science
Foundation operated by Associated Universities, Inc.  This material is
based upon work supported by the National Science Foundation Graduate
Research Fellowship under Grant No. DDGE-1315231.  Support for this
work was provided by the NSF through the Grote Reber Fellowship
Program administered by Associated Universities, Inc./National Radio
Astronomy Observatory.  THJ acknowledges support from the National
Research Foundation (South Africa).  This research has made use of the
NASA/IPAC Infrared Science Archive, which is funded by the National
Aeronautics and Space Administration and operated by the California
Institute of Technology.
We thank the anonymous referee for a careful reading and detailed suggestions
for clarifying our manuscript.\\
\end{acknowledgments}

\facilities{Gaia, IRSA, MeerKAT, NED}
\vfil\eject

\bibliographystyle{aasjournal}
\bibliography{arXiv}

\end{document}